\documentclass[
aps,prb,
twocolumn,
groupedaddress]{revtex4}
\input{epsf}
\usepackage{epsfig} 
\usepackage{amsfonts}
\usepackage{pifont}
\usepackage{epstopdf}
\usepackage{epsfig}
\usepackage{amsfonts}

\def\mts2f{${\cal M}({\cal T}^{*(2F)})$}
\def\mtsa4{${\cal M}({\cal T}^{*(A_4)})$}

\def\ts2f{${\cal T}^{*(2F)}$}

\def\cstsa4{${\cal C}^s_{{\cal T}^{*(A_4)}}$}
\def\tsa4{${\cal T}^{*(A_4)}$}

\def\es{$\mathbb{E}_\perp$}

\def\d*six{$D^*_6$}

\usepackage{times}
\usepackage{pifont}

\def\zfo{\ding{193}}
\def\dfo{\ding{194}}
\def\ffo{\ding{196}}

\begin{document}

\title{ 
Thick atomic layers of maximum density as 
bulk terminations of quasicrystals 
}

\author{Zorka Papadopolos}
\email[]{zorka.papadopolos@uni-tuebingen.de}

\affiliation{Institut f\"ur Theoretische Physik, 
Univ. T\"ubingen, D-72076 T\"ubingen, Germany}

\author{Gerald Kasner}
\affiliation{Institut f\"ur Theoretische Physik, 
Univ. Magdeburg, PF 4120, D-39016
  Magdeburg, Germany}

\date{\today}

\begin{abstract}
The clean surfaces of quasicrystals, 
orthogonal to the directions of the main symmetry axes, 
have a terrace-like appearance. 
We extend the Bravais' rule for crystals 
to quasicrystals, allowing that instead of a
\emph{ single} atomic plane 
a \emph{ layer} of atomic planes may form a bulk 
termination.

\end{abstract}


\maketitle

We consider clean surfaces of quasicrystals
by comparing their STM (scanning tunneling microscopy)
and SEI (secondary electron imaging) images 
to the bulk terminations in the deterministic atomic 
models. 
A bulk model is defined as an ideal quasiperiodic 
tiling with atomic decoration on a finite number of 
tiles. In Section~\ref{sec:decagonal}, 
an STM image of a decagonal quasicrystal is compared 
to the Burkov model\cite{Bur93} \mtsa4, 
based on a decagonal tiling \tsa4, 
see Ref.\cite{Baake90}. 
In section~\ref{sec:ico}, STM and SEI images of an 
icosahedral quasicrystal are compared to the 
model~\cite{Aper97}
\mts2f, based on icosahedral tiling  
\ts2f, see Refs~\cite{
ts2f,
PHK}. 
The clean surfaces, that we study, are considered not 
to be reconstructed.

In our previous work on  terrace-like clean surfaces 
of icosahedral quasicrystals we tried to define 
the 5fold terminations over the most dense layers of 
Bergman polytopes~\cite{KPKB},
which are the icosahedral motifs on a 
quasilattice \ts2f.  
Later, we noticed that the
5fold terminations should better be defined by the 
0.48~{\AA} thin, plane-like atomic layers, all of equal, 
maximum density~\cite{LGPK2}. 
In Ref.~\cite{PP2} we adopted  Bravais' rule
of maximum density, 
allowing that also the densities of thin 
($< 0.6$~{\AA} broad), plane-like atomic layers in 
quasicrystals should be treated as the planes. 
The idea we developed in Ref. \cite{PP2}, and gave 
a possible explanation of the experimental 
fact, that although the 2fold planes in the model 
\mts2f\ are more dense than the 5fold planes, 
the 5fold surfaces are the most stable in 
icosahedral quasicrystals \cite{Shen00}.
In Ref.~\cite{PP2}, in which we used the ``thin'' 
layer concept in order to determine the terminations
in all principle symmetry directions, we announced the 
``thick'' layer concept as well, that we elaborate in 
the present manuscript.

\section{Surfaces of decagonal quasicrystals
\label{sec:decagonal}}

\begin{figure}[]
\includegraphics[width=0.27\textwidth]
{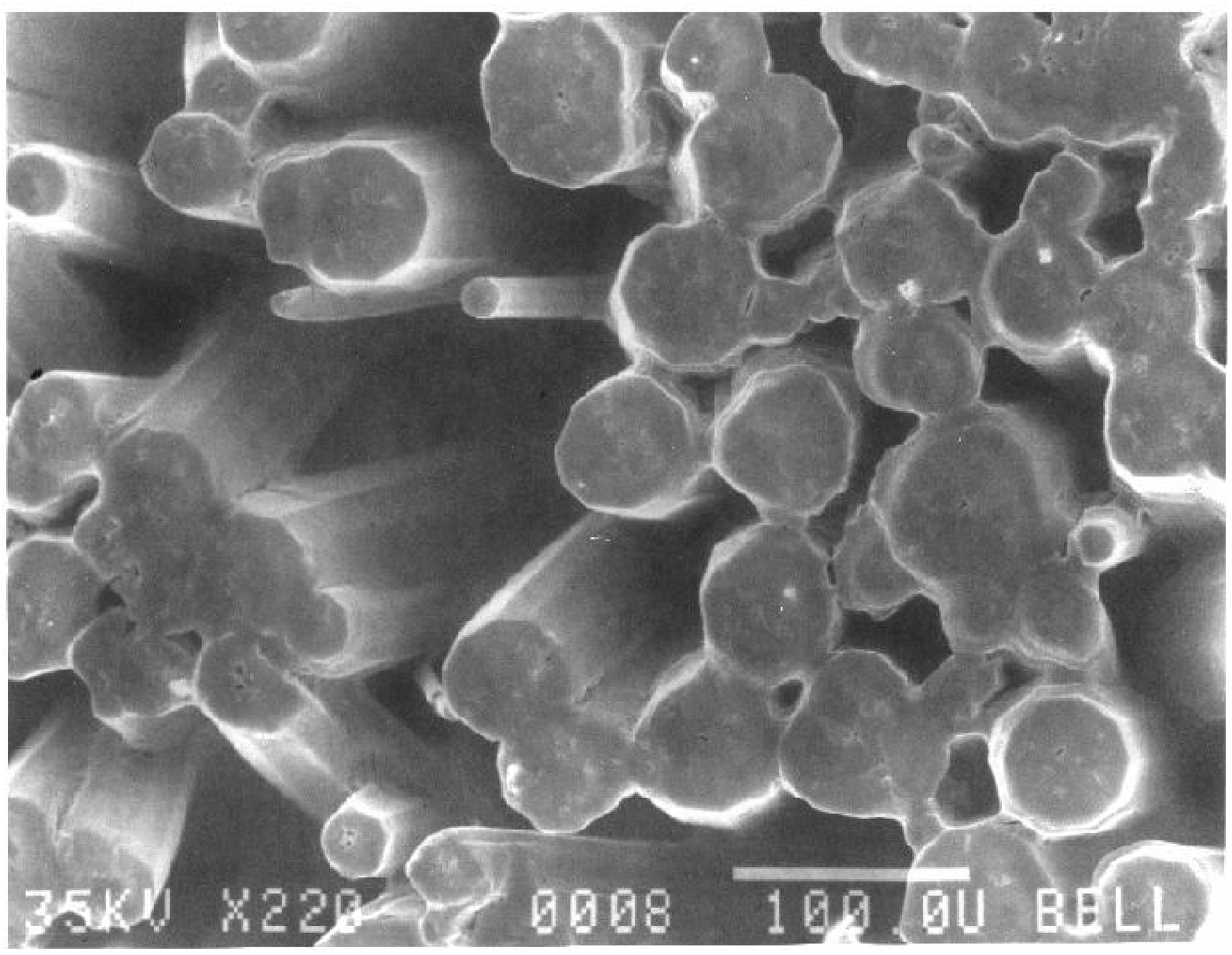}
\includegraphics*[width=0.30\textwidth]
{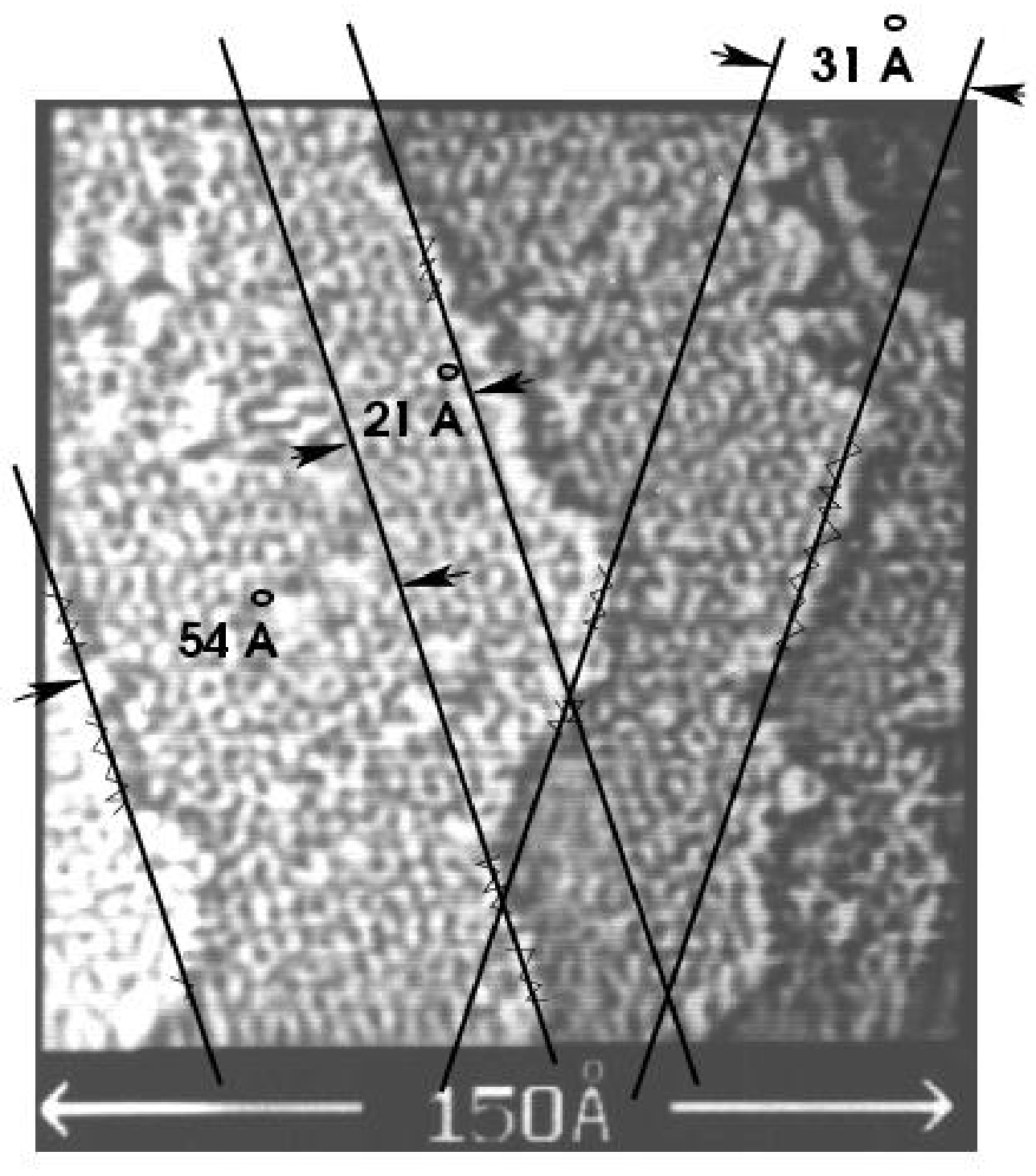}
\includegraphics*[width=0.40\textwidth]
{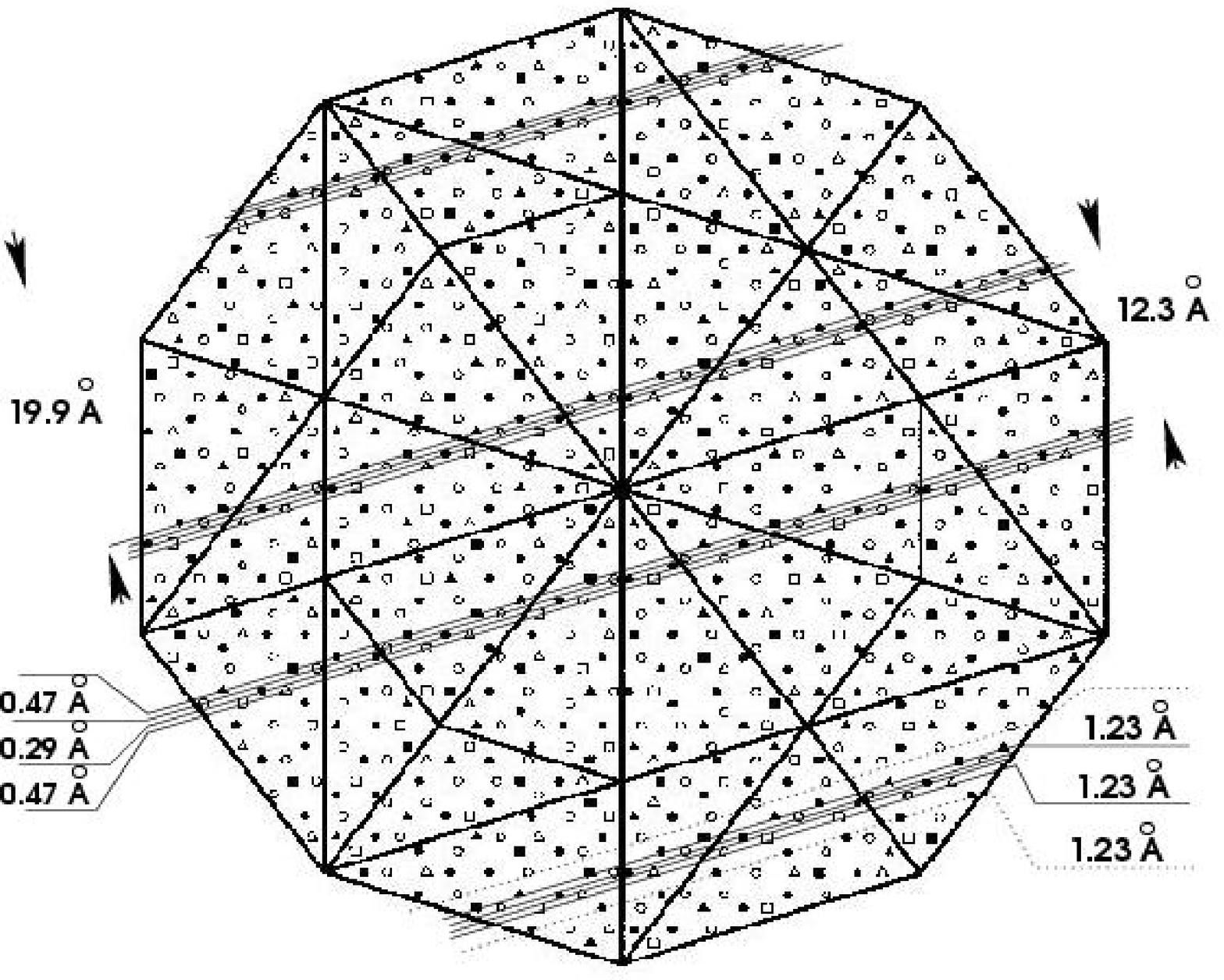}
 \caption{
 (a)~(top) Long, thin decagonal prisms~\cite{K0} 
           of the d-AlCuCo;
 (b)~(middle) Estimated positions of 2fold surfaces 
        on an STM image of the 10fold terrace-like 
         surface of d-AlCuCo;
 (c)~(bottom) 2fold bulk terminations in \mtsa4\  
         marked on a 10fold one~\cite{Kram95}. 
      Note the $1.23$~{\AA} gaps on both sides of 
      a termination. 
\label{fig:deca}
}
\end{figure}

Decagonal Al$_{65}$Cu$_{15}$Co$_{20}$ (d-AlCuCo) is 
periodic in $z$-direction~\cite{Kor90}, 
with periodicity t=4.13~{\AA}. The  $z$-direction is
orthogonal to the quasiperiodic decagonal $x-y$ plane. 
The phase crystallizes in a shape of long thin 
decagonal prisms, with the 2fold surfaces considerably 
larger than the 10-fold surfaces, 
see Fig.~\ref{fig:deca}(a).
%
%
%
The size of the surface area is evidently a parameter 
of it's stability. 
In Fig.~\ref{fig:deca}(b), over the image of the 10fold 
surface \cite{Kor90}, we determine the positions of 
possible 2fold surfaces, orthogonal to the 10fold 
surface. 
In the model \mtsa4\ (t=4.18~{\AA})~\cite{Bur93} we 
investigate the densities of the ``thin'', 2fold layers 
containing 2 atomic planes. 
Among these, the most dense one is a $0.47$~{\AA}-layer 
of the density $\rho_{2}^{0.47A}=0.124$~{\AA}$^{-2}$.
Comparing it to the much smaller 10fold surface, 
which is of the density $\rho_{10}=0.146$~{\AA}$^{-2}$, 
we conclude that the $0.47$~{\AA}-layer can not 
represent the 2fold termination. But, on some positions
in the bulk these layers appear in 
pairs, $0.29$~{\AA} apart. Such, a rather ``thick'' 
$1.23$~{\AA}-layer of 4 planes on mutual
small distances (see Fig.~\ref{fig:deca}(c)) is a 
candidate for a terminating layer. These layers appear 
on distances mutually scaled by the factor 
$\tau=(1+\sqrt{5})/2$ ($12.3$~{\AA}, $19.9$~{\AA}, 
$32.2$~{\AA}, $52.1$~{\AA}\ldots, 
as in Fig.~\ref{fig:deca}(c)), in excellent agreement 
with those found on the STM image ($21$~{\AA}, 
$31$~{\AA}, $54$~{\AA}, see Fig.~\ref{fig:deca}(b)).

\section{ Surfaces of icosahedral quasicrystals
\label{sec:ico}}

A feature of the surface of \mbox{i-AlPdMn}, not
accounted for by the ``thin'' layer analysis \cite{PP2}, 
is that not all types of maximally dense layers appear 
as surfaces:
for example, $(q,b)$ layers~\cite{classes}, 
$0.48$~{\AA} apart, are seen in 5fold
surfaces but equally dense $(b,q)$ layers, 
also $0.48$~{\AA} apart, 
are not. If both kind of layers were possible 
terminations, the sequence of much 
shorter terrace heights, than observed, could appear. 
%
%
\begin{figure}[]
\includegraphics*[width=0.48\textwidth]
{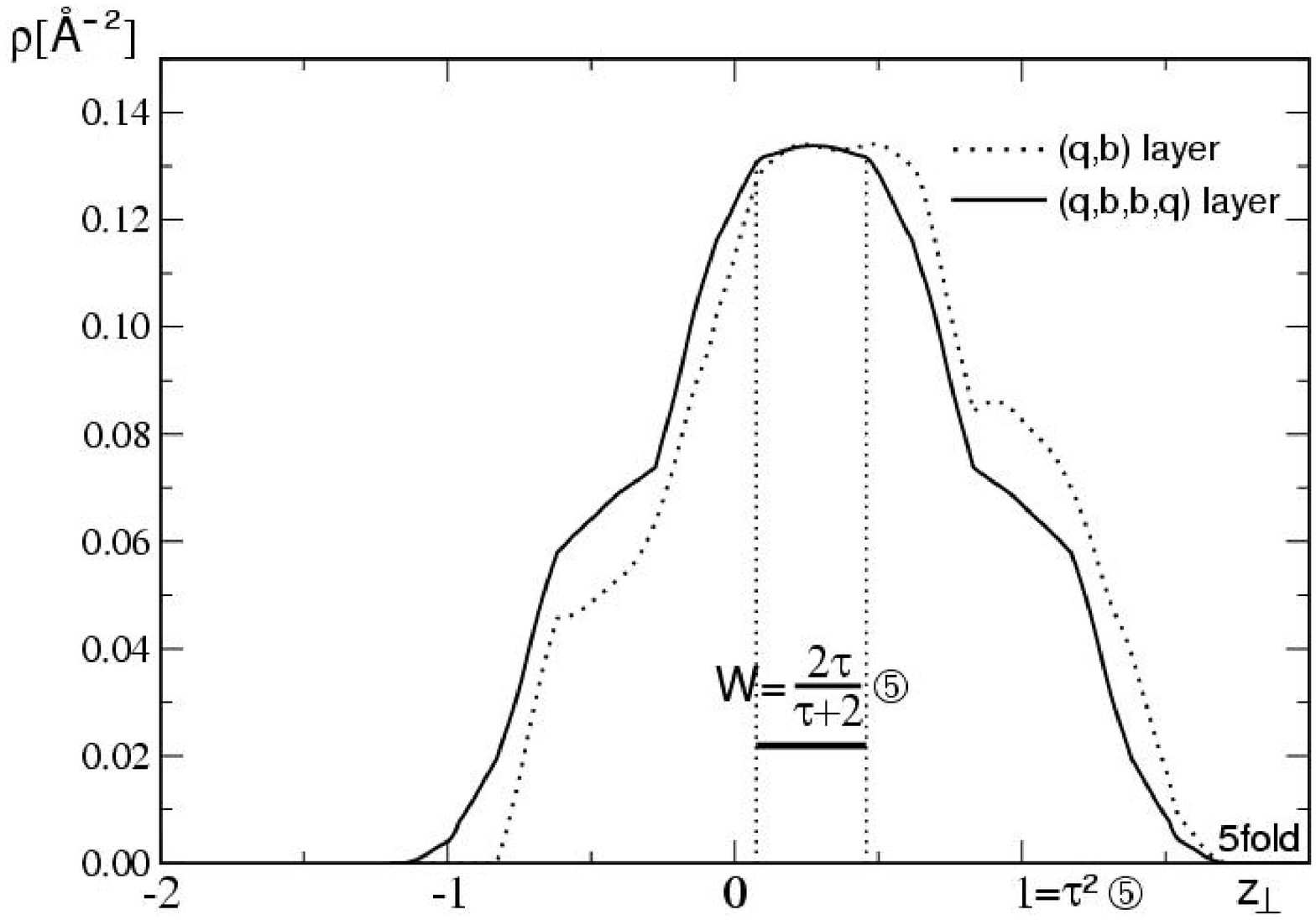}
\includegraphics*[width=0.48\textwidth]
{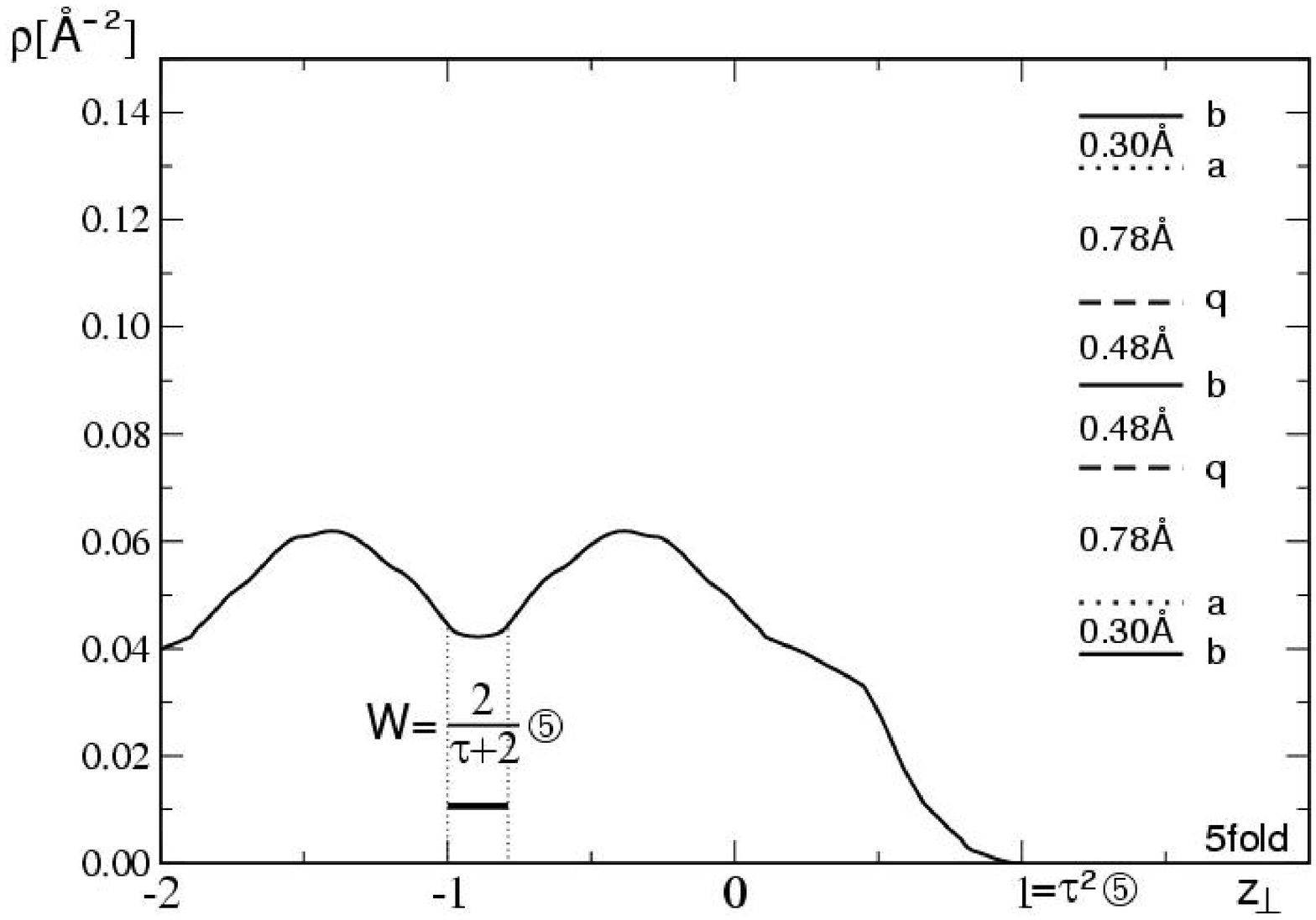}
\caption{
(a)~(top)
Density graph $\rho_{5f}(z_{\perp})$ of the ``thick'' 
5fold layers $(q,b,b,q)$ (full line). The plateau 
(maximum density) of the graph defines the terminations. 
It is compared to the density graph of the $0.48$~{\AA} 
thin, plane-like layer $(q,b)$ (dotted line).
The symbol \ffo\ is the standard distance along 
a 5fold axis $z_{\perp}$ in the coding space \es.
(b)~(bottom) Density graph of the 
5fold layers $(b,a,q,b,q,a,b)$, with spacings as in the
image. 
The bottom of the 
cavity
defines a sequence of the 
{\em minimum density} layers in \mts2f, situated above 
a {\em subsequence} of the
 5fold terminations.  
\label{fig:5f-thick-term}
}
\end{figure}
If one chooses to define a
termination incorporating the neighboring planes too,
as we did in decagonal case 
(in Section~\ref{sec:decagonal}),
one could introduce a ``thick'' layer as a bundle of
high density planes (or thin, plane-like layers).  
A 5fold termination can be considered to
be a ``thick'' layer consisting of a 
$(q,b)$  layer and a $(b,q)$ layer, 
each with the spacing $0.48$~{\AA}.
Such a layer contains 4 planes with
spacings:
$q$-plane, $0.48$~{\AA}, $b$-plane, $1.56$~{\AA}, 
$b$-plane, $0.48$~{\AA}, $q$-plane.  
For a bundle we define an effective
(averaged) density of within contained ``thin'' 
layers/planes

\vspace{0.3cm}
\noindent
{ \small
$\rho_{5f}(z_{\perp})=({\rho}_{q_1}(z_{\perp})+
{\rho}_{b_1}(z_{\perp}))/2 +
({\rho}_{q_2}(z_{\perp})+
{\rho}_{b_2}(z_{\perp}))/2$
}.

\vspace{0.3cm}
\noindent
As we see in Fig.~\ref{fig:5f-thick-term}(a), 
whereas for the thin-layer concept the width of the
support of the plateau is approximately 
$\frac{2{\tau}^2 }{\tau+2}$\ffo\ broad,  
and consequently~\cite{PP2} encodes the 
Fibonacci sequence of terrace heights $S=4.08$~{\AA} 
and $L=\tau S=6.60$~{\AA} 
($\tau=(1+\sqrt{5})/2$),
in the thick-layer concept the width is 
exactly
$\frac{2{\tau} }{\tau+2}$\ffo\  
(see  Fig.~\ref{fig:5f-thick-term}(a)) 
and encodes the Fibonacci sequence of by factor $\tau$ 
larger terrace heights, i.e. $L=6.60$~{\AA} and 
$L+ S=\tau L=10.68$~{\AA}. 
Whereas on the clean surfaces, obtained at lower 
annealing temperature, even the terrace height 
$\tau^{-1}S=2.52$~{\AA} appears~\cite{Shen99}, 
on the 
surfaces obtained at the highest 
annealing temperature, on the contrary, 
a terrace height $L+ S$ appears   
(see Fig.~1 in Ref.~\cite{KPKB}).

The height of the plateau  of the graph 
$\rho_{5f}$ (see Fig.~\ref{fig:5f-thick-term}(a)) 
defines the densities of the ``thick'' layer 
terminations to be $0.134$~{\AA}$^{-2}$, 
see table~\ref{tab:densities}.
The 5fold layers intertwining the terminations are 
of densities not higher than 
$0.072$~{\AA}$^{-2}(\ll 0.134$~{\AA}$^{-2}$). 
It is also a fact, that the density graphs of the 
``thin'' and the ``thick'' layers have a strong overlap,
see Fig.~\ref{fig:5f-thick-term}(a).
Hence, almost any  $(q,b)$ ``thin'' layer termination 
occurs within such a $(q,b,b,q)$ ``thick'' layer 
termination. 
The same holds true for any icosahedral quasicrystal 
described by the \mts2f\ model.

Above each termination, there is a $2.04$~{\AA} gap,
if we dare to neglect an $a$-plane of a density
smaller than $0.013$~{\AA}$^{-2}$. 
But, if each gap of $2.04$~{\AA} 
in the model \mts2f\ would be declared as 
a criterion of a termination to appear below it, 
as in Ref.~\cite{Y04}, the $2.04$~{\AA} terrace heights, 
that were not observed, should appear as well. 
However in the model \mts2f\ there is a 
low density 5fold layer $(b,a,q,b,q,a,b)$ 
(see Fig.~\ref{fig:5f-thick-term}(b)), 
$4.08$~{\AA} broad. 
The width of the
cavity 
on the density graph of these layers, 
$W=(2/(\tau +2))$\ffo, encodes a Fibonacci
sequence with the intervals 
$\tau L=10.68$~{\AA} and  $\tau^2 L=17.28$~{\AA}. 
These are the {\em minimum density} layers
of equal, $0.041$~{\AA}$^{-2}$ density, 
placed in the model over a {\em subsequence} of 
the terminations.
Hence, the minimum density layer sequence 
alone can not define
the terminations, because 
it does not reproduce 
the pairs of large terraces $6.60$~{\AA} 
apart, which were frequently observed.

%
%
\begin{figure}[]
\includegraphics*[width=0.48\textwidth]
{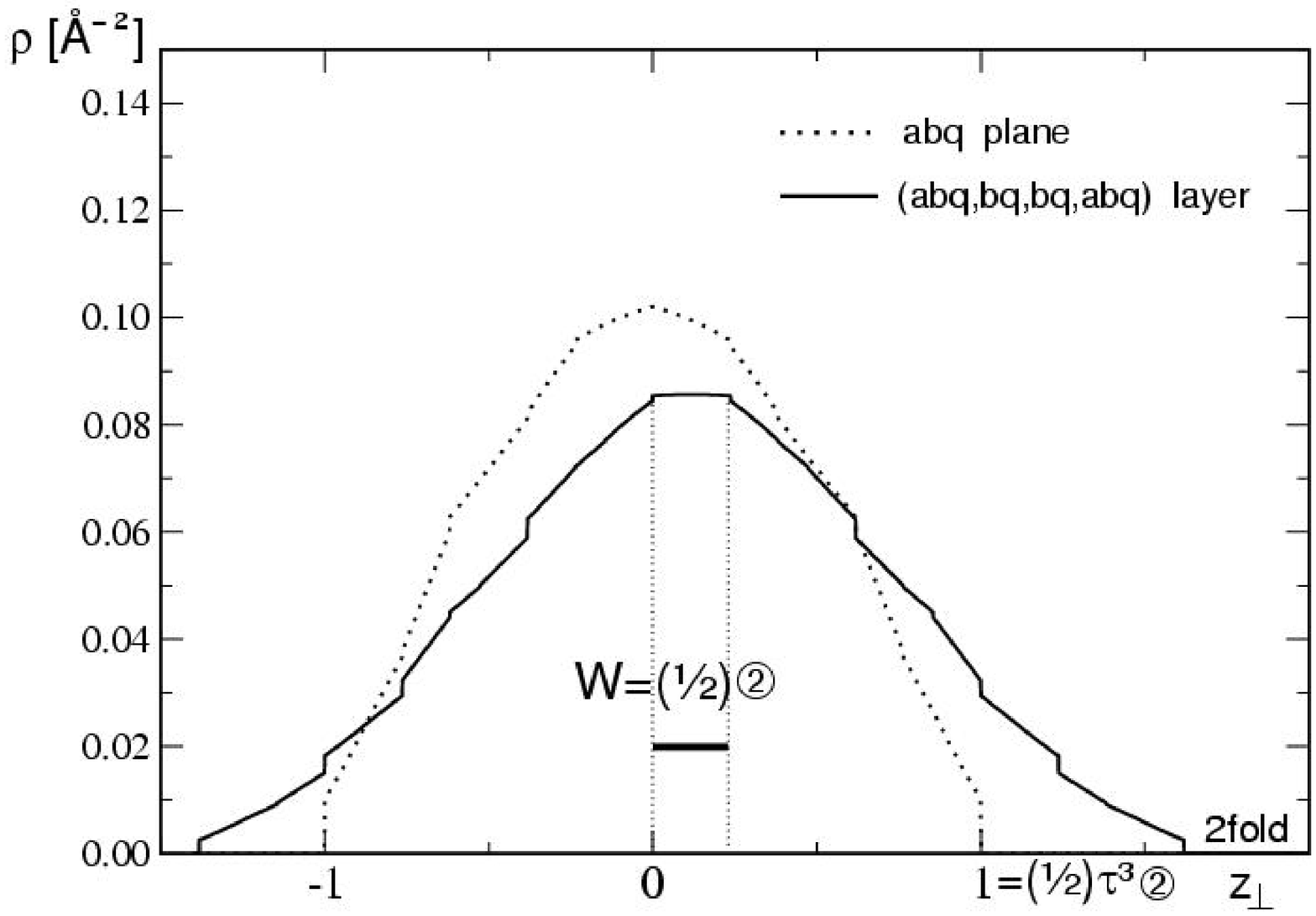}
\includegraphics*[width=0.48\textwidth]
{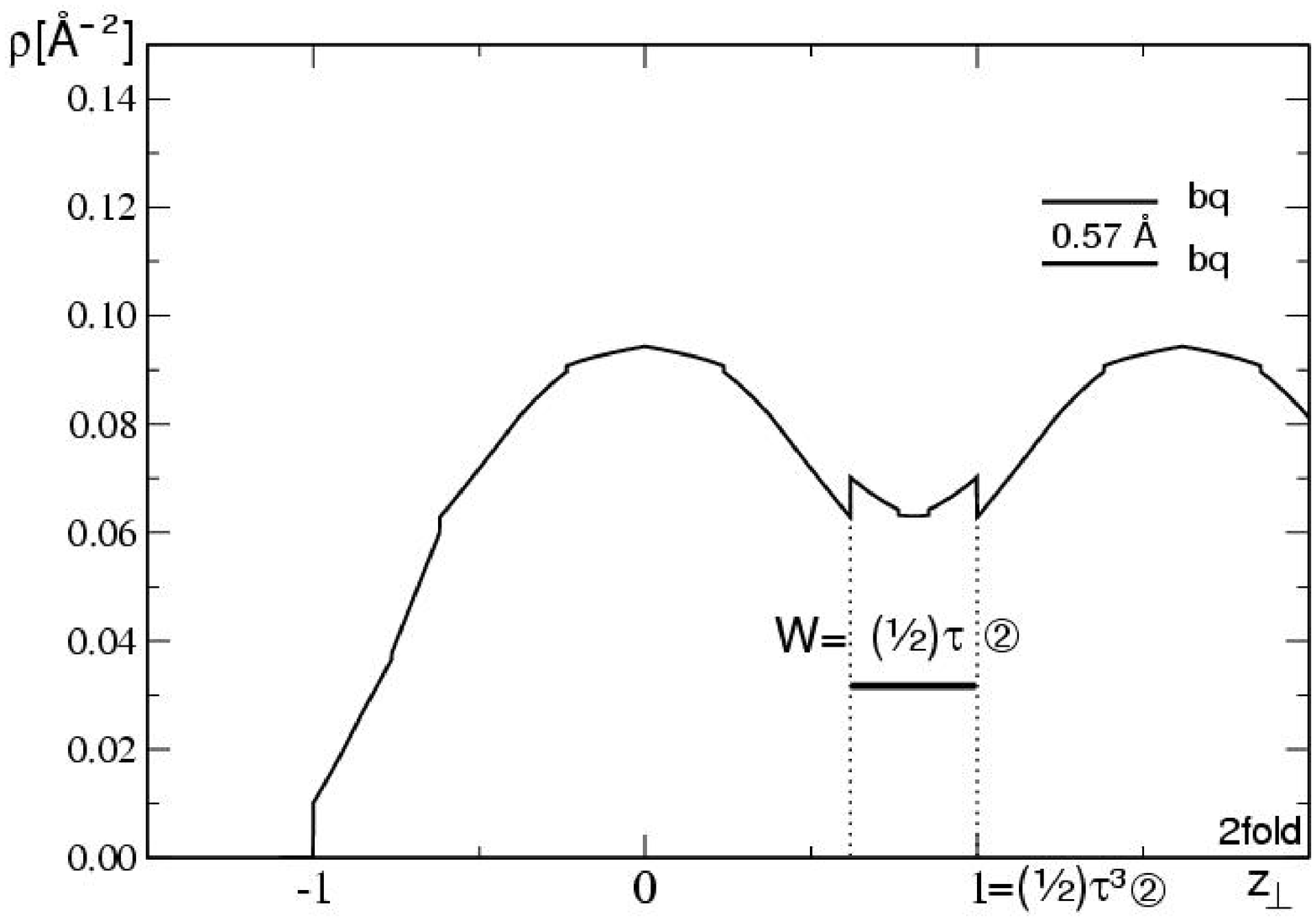}
  \caption{
(a)~(top) 
Density graph $\rho_{2f}(z_{\perp})$ 
of the ``thick'' 2fold
layers (full line). The plateau (maximum density) 
of the graph defines the terminations. It is compared 
to the ``thin'' layer termination (dotted line), 
which is a single, $abq$-plane termination. 
The symbol \zfo\ is the standard distance 
along a 2fold axis $z_{\perp}$ in the coding space
\es. 
(b)~(bottom) 
Density graph of the 
2fold layer $(bq,bq)$. The cavity
defines a sequence of the {\em  minimum density} layers
in \mts2f,
of which a {\em subsequence} is situated above 
{\em all} the 2fold terminations.  
\label{fig:2f-thick-term}
}
\end{figure}
In the case of 2fold surfaces, we may replace a single 
dense 2fold terminating atomic
$abq$-plane~\cite{PP2} by a layer of 4 atomic planes
with spacings:
$abq$-plane, $1.48$~{\AA}, $bq$-plane, $0.92$~{\AA},
$bq$-plane, $1.48$~{\AA}, $abq$-plane. 
For a bundle we define an effective
(averaged) density of planes

\noindent
\vspace{0.3cm}
\noindent
{\small$\rho_{2f}(z_{\perp})=
(1/4)[{\rho}_{abq_1}(z_{\perp})+
 {\rho}_{bq_1}(z_{\perp})+
 {\rho}_{bq_2}(z_{\perp})+
 {\rho}_{abq_2}(z_{\perp})$]},

\vspace{0.3cm}
\noindent
For this ``thick'' 2fold layer
the peak of $\rho_{2f}(z_{\perp})$ is a perfectly 
flat plateau, see Fig.~\ref{fig:2f-thick-term}(a).
The height of the plateau defines the effective
densities of terminations to be $0.086$~{\AA}$^{-2}$, 
see Table~\ref{tab:densities}. 
The support of the width of the plateau equals  
$W=(1/2)$\zfo\ and encodes the 
Fibonacci sequence of 2fold terminations with the 
terrace heights $S=6.3$~{\AA} and 
$L=\tau S=10.2$~{\AA}. 
The standard distances along 5, 2 and 3fold axes 
in icosahedral structures are 
\ffo~/$\sqrt{\tau +2}={}$~\zfo~/2$(=1/\sqrt{2(\tau +2)})
={}$~\dfo~$/\sqrt{3}$, where $\tau=(1+\sqrt{5})/2$.
The heights of the larger 2fold terraces were 
measured to be
$S=
6.2$~{\AA} and $L=9.5$~{\AA}
(see Fig.~3(a) and 3(c) in Ref.~\cite{PP2}),
in good agreement with the predicted values.
The 2fold layers intertwining the terminations are 
of densities not higher than 
$0.079$~{\AA}$^{-2}(<0.086$~{\AA}$^{-2}$). 
The small 2fold terraces, or rather the 
pits within the big terraces  
(see Fig.~3(a) and 3(c) in Ref. \cite{PP2})
may be explained by the comparatively large distances 
between the atomic planes inside of the ``thick''
terminating layer ($1.5$~{\AA}, $2.4$~{\AA} and 
$3.9$~{\AA}).
These excellently reproduce the measured values
($2.4$~{\AA} and $3.6$~{\AA}), 
see Fig.3(a) and 3(c) in Ref. \cite{PP2}.

In the model \mts2f\ there is a low density 
2fold layer 
$(bq,bq)$: 
$0.92$~{\AA} gap, $bq$-plane, $0.57$~{\AA},
$bq$-plane, $0.92$~{\AA} gap
(see Fig.~\ref{fig:2f-thick-term}(b)). 
The width of the
cavity
on the graph of these layers, 
$W=(\tau/2)$\zfo, encodes a Fibonacci
sequence with the intervals 
$\tau^{-1} S=3.9$~{\AA} and $S=6.3$~{\AA}. 
These are the {\em minimum density} layers
of almost equal density, 
somewhat above $0.063$~{\AA}$^{-2}$.
A member of a {\em subsequence} of these layers 
is placed over  {\em each} 2fold termination.
Nevertheless, in the 2fold case the 
minimum density layer sequence {\em alone} 
can not define the terminations, because it predicts 
by $\tau^{-1}$ shorter terrace heights between the 
large 2fold terminations, than observed.

The 3fold terminations could also be modeled
as ``thick'' layers of atomic planes in \mts2f, 
see table~\ref{tab:densities}.
But, inspecting the intertwining 3fold layers,
we see that these are of the densities {\em comparable} 
to the ``terminating'' ones.  
We also know that the 3fold surfaces facet readily 
\cite{Shen00}, 
and some correlated STM measurements, 
(as those in Fig.3(a) and 3(c) of Ref. \cite{PP2})
for the 3fold surfaces do not exist so far.
%
%
\begin{table}[h]
\begin{center}
\caption{ 
Relative and absolute densities of the planes
and layer terminations orthogonal to 5, 2 and 3fold 
symmetry axes in \mts2f\  of \mbox{i-AlPdMn}.
The corresponding data for \mbox{i-AlCuFe} are similar. 
\label{tab:densities}
}
\begin{tabular}{l|lll}
\hline
   & 5fold & 2fold & 3fold\\
\hline
Densest planes (abs.) & 0.086\,{\AA}$^{-2}$&
0.101\,{\AA}$^{-2}$&0.066\,{\AA}$^{-2}$ \\
Densest ``thin'' layers (abs.) &0.133\,{\AA}$^{-2}$&
0.101\,{\AA}$^{-2}$&0.066\,{\AA}$^{-2}$\\
Densest ``thin'' layers (rel.) & 1 & 0.76 & 0.50\\
Densest ``thick'' layers (abs.) &0.134\,{\AA}$^{-2}$&
0.086\,{\AA}$^{-2}$&0.058\,{\AA}$^{-2}$\\
Densest ``thick'' layers (rel.) & 1 & 0.64 & 0.44\\
\hline
\end{tabular}
\end{center}
\end{table}

On the STM measurements it is in general 
hard to judge whether ``thin'' layer or ``thick''  
layer terminations best model the physical surfaces.  
However, in the 5fold case, the ``thick'' layer concept 
removes the contradiction with respect to the 
Bravais' rule, that some, equally dense layers 
do not appear on the surfaces, see Ref.~\cite{PP2}.
In the 2fold case the ``thick'' layer concept is 
evidently better, it treats differently
the large terraces compared to the small pits inside. 
That the effective densities of the 2fold terminations 
are somewhat lower than the densities of some single 
2fold planes is not contradictory, because these are 
included in the most dense layers.

\begin{figure}
\includegraphics[width=0.465\textwidth]
{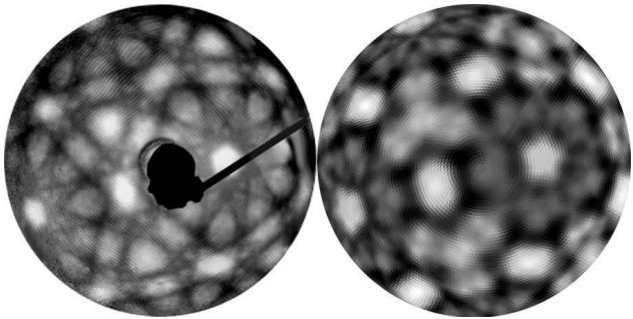}
  \caption{
   (a)~(left) Secondary-electron pattern obtained from 
       the pentagonal surface of a single icosahedral 
       Al$_{70}$Pd$_{20}$Mn$_{10}$ quasicrystal. 
       The center of the pattern is obscured by the 
       shadow of the electron gun used for the 
       excitation. The edge of the screen corresponds 
       to~\cite{B04} $\theta=52^\circ$.  
(b)~(right) Calculated secondary-electron pattern based 
    on the single scattering approximation of electrons 
    using model~\cite{E03} \mts2f.
\label{fig:compSEI}
}
\end{figure}

Fig.~\ref{fig:compSEI}(a)
shows the secondary-electron pattern 
\cite{B04}
obtained from the clean pentagonal surface of a 
quasicrystalline 
Al$_{70}$Pd$_{20}$Mn$_{10}$ 
sample~\cite{K0}. 
%
%
Secondary-electron images 
(SEI) represent an orthogonal 
projection to the sphere of the symmetry directions 
below a near-surface region 
of the sample. 
Apart from the icosahedral symmetry of the pattern, 
some groups of bright patches are seen to lie 
within bands, similar to Kikuchi bands 
\cite{B04} connecting the 
2fold-, 3fold-, and 5fold-symmetry directions.
In case of crystals these
bands are a direct consequence of well-defined dense 
{\it planes} 
of atoms.
The bands observed by a quasicrystal 
should be also a consequence of dense planes or  
plane-like layers within the quasicrystal, which, 
according to the secondary-electron pattern, 
lie  perpendicular to the principal directions of the 
icosahedron. 
Hence, the pattern in Fig.~\ref{fig:compSEI}(a) 
carries an
information of the long-range order. 
We note that there are bands 
perpendicular to 2fold- and 5fold-symmetry directions, 
but not to 3fold-symmetry directions.

A quantum mechanical single-site scattering calculation 
\cite{B04} is a faithful representation of the SEI 
pattern because it accounts for the wave nature of the 
secondary electrons. 
Fig.~\ref{fig:compSEI}(b) illustrates the results of 
the calculation~\cite{E03} using the coordinate set 
of the model \mts2f. This approximation overestimates 
the scattering intensity along chains of atoms 
\cite{B04}
but suffices for 
purposes, since our interest is 
mainly in the presence or absence of bands on the 
screen. 
As in case of crystals \cite{B04} the band width 
is inversely proportional to the spacing of 
crystallographic planes, 
and the band width is related to the distance 
of interatomic planes 
\cite{B04}.
The observed band widths reveal that the 
inter-planar distances of the highly dense 
2fold planes are broader than these of the 
5fold planes/plane-like layers 
by a factor  
1.6.
In 5fold case, from the model we predict the distance  
$d_5=\tau/(\tau+2)$\ffo$=2.04$~{\AA}, which is the 
distance between the highly dense $(q,b)$ and the 
$(b,q)$ plane-like layers in the terminating 
thick layer $(q,b,b,q)$: 
$2.04$~{\AA}$=0.48$~{\AA}$+1.56$~{\AA}, 
see also Ref.~\cite{Z04}.
In the 2fold case, between the highly dense planes 
in the above defined ``thick'' layer, appears
once the distance of 0.92~{\AA} and twice the distance 
of 1.48~{\AA}. 
Hence, an average distance between the highly dense 
planes is 
$d_2=\frac{\tau}{3\sqrt{\tau+2}}$\zfo$=1.29$~{\AA}. 
And their ratio is $d_5/d_2=3/\sqrt{\tau+2}
\approx 1.6$. 

The SEI method does not determine the bulk termination, 
(it is testing the bulk circa 30~{\AA} below 
the surface). However, 
in the case of the ordinary 
crystals, the Kikuchi bands are related to the most 
dense atomic planes, and we show that the same holds 
true in the case of quasicrystals as well. 
Hence we may claim that SEI images are supporting the 
thick atomic layers of the high effective 
density to be the bulk terminations, if the Bravais' 
rule should be valid in quasicrystals as well.  
The not existing 3fold Kikuchi bands  are also 
supporting 
the model \mts2f, in which we find that the 
atoms collected by the 3fold  planes are almost 
uniformly distributed among these, without the 
notable repetitive layers of higher densities.

\end{document}